\documentclass[12pt,a4paper]{article}
\usepackage{epsfig,amsmath,amssymb,subfigure,placeins}
\usepackage[colorlinks=true,linktocpage=true,linkcolor=blue,citecolor=blue]{hyperref}
\usepackage{color}
\newcommand{\old}[1]{}

\newcommand{\be}{\begin{equation}}
\newcommand{\ee}{\end{equation}}
\newcommand{\ba}{\begin{eqnarray}}
\newcommand{\ea}{\end{eqnarray}}
\newcommand{\bi}{\begin{itemize}}
\newcommand{\ei}{\end{itemize}}

\begin{document}
\begin{flushright}
{\normalsize
}
\end{flushright}
\vskip 0.1in
\begin{center}
{\large{\bf Velocity-induced Heavy Quarkonium Dissociation using the 
gauge-gravity correspondence}}
\end{center}
\vskip 0.1in
\begin{center}
Binoy Krishna Patra, Himanshu Khanchandani and Lata Thakur \\ 
{{\it Department of Physics, Indian Institute of 
Technology Roorkee, Roorkee 247 667, India} }
\end{center}
\vskip 0.01in
\addtolength{\baselineskip}{0.4\baselineskip} 


\begin{abstract}
Using the gauge-gravity duality we have obtained the potential 
between a heavy quark and an antiquark pair, which is
moving perpendicular to the direction of orientation, in a strongly-coupled 
supersymmetric hot plasma. For the purpose
we work on a metric in the gravity side, {\em viz.} OKS-BH geometry, whose
dual in the gauge theory side runs with the energy and hence proves to be a
better background for thermal QCD. 
The potential obtained has confining term both in vacuum and in medium,
in addition to the Coulomb term alone, usually reported in the literature.
As the velocity of the pair is increased the screening of the potential 
gets weakened, which may be understood by the decrease of effective 
temperature with the increase of velocity. The crucial observation of
our work is that beyond a critical separation of the heavy quark pair, the 
potential develops an imaginary part which
is nowadays understood to be the main source of dissociation.
The imaginary part is found to vanish at small $r$, thus agrees with the 
perturbative result. Finally we have estimated the thermal width
for the ground and first excited states and found that
non-zero rapidities lead to an increase of thermal width.
This implies that the moving quarkonia dissociate easier than the
static ones, which agrees with other calculations. However, the 
width in our case is larger than other calculations due to the presence 
of confining terms.
\end{abstract}
\section{Introduction}
The in-medium behaviour of heavy quark bound states is used to probe the 
state of matter in quantum chromodynamics (QCD), where the screening of 
the potential results in the suppression of the yields of heavy quarkonium 
states in relativistic heavy-ion 
collisions~\cite{Matsui:1986dk}. 
The heavy quarkonium bound states treated nonrelativistically
possess the well separated energy scales, {\em viz.}
the heavy quark mass ($m_Q)$ $\gg$ the momentum transfer ($1/r \propto  
m_Q \alpha_s$) $\gg$ the binding energy ($E \propto  m_Q \alpha_s^2$)
\cite{Brambilla:2004jw}. 
In addition, the (thermal) medium introduces 
scales to the previous ones, {\em viz.} T, the Debye mass, $m_D$ ($gT$) etc.
which are also separated (T $\gg$ $m_D$) 
in weak coupling regime ($g<1$)\cite{Brambilla:2008cx}. The above hierarchies 
facilitate to develop the sequence of effective theories {\em namely} 
NRQCD, pNRQCD etc. from the underlying theory of strong interaction (QCD)
after integrating out the successive energy scales for both 
T=0 and T $\ne$ 0\cite{pNRQCD1}. Now if the $Q \bar Q$ bound state 
moves with respect to the medium, the synthesis of effective theories
becomes more complicated due to the additional scales associated 
with the motion of the pair. However, EFT have recently explored 
the in-medium modifications of heavy quarkonium states moving through 
a medium for two plausible situations: 
$m_Q \gg 1/r \gg T \gg E \gg m_D$ and $m_Q \gg T \gg 1/r \,, m_D  \gg E$
~\cite{Escobedo:PRD2013}, which are relevant for moderate temperatures
and for studying dissociation, respectively. They found that 
the width decreases with the velocity for the former situation whereas for 
the
latter regime the width increases monotonically with the velocity.

The aforesaid hierarchies assume the weak coupling limit and its 
identification with the relative velocity. 
Therefore one needs the framework in strong coupling
regime, where the potential can be extracted from the Euclidean correlators 
calculated in the lattice but the limited sets of data
in addition to the technical difficulties of lattice calculations,
limits the reliability of the results. Thus some complementary methods 
for strong coupling are desirable, where the AdS/CFT conjecture 
provides such alternative~\cite{Maldacena:1998adth,Witten:1998,Gubser:1998} to
calculate the potential for a class of non-abelian thermal gauge field theories.
The expectation value of a particular time-like Wilson loop defines the 
potential between a static quark and antiquark at finite temperature.
This is where the AdS/CFT correspondence makes the calculation easier
by mapping the 
evaluation of a Wilson loop in a hot strongly interacting gauge theory 
plasma onto the much simpler problem of finding the extremal area of 
a classical string world sheet in a black hole background.
The first calculation was done by Maldacena~\cite{Maldacena:1998im}
for $ {\cal N} = 4 $ SYM for T=0 and was later extended to finite temperature
in~\cite{Brandhuber:1998bs,Rey:1998bq}.

The physics of quarkonium dissociation in a medium has undergone 
refinements over the last decade~\cite{Laine:2008cf,BKP:2000,BKP:2001,
BKP:2002,BKP:2004. Physically a resonance is disolved into a medium 
by the broadening of its peak due 
to its interaction with the constituents of the medium. Earlier a quarkonium state
was thought to be dissociated when the Debye screening becomes so strong that it inhibits 
the formation of bound states but nowadays a quarkonium is understood to be dissociated 
at a lower temperature~\cite{Laine:2006ns, otherrefsImV} even though its binding energy is 
nonvanishing rather is overtaken by the  Landau-damping
induced thermal width, which may be obtained from the imaginary part of the 
potential ~\cite{Laine:2006ns,Brambilla:2008cx,Laine:2008cf}.} 
There are mainly two processes in QCD, which contribute to the
thermal width: the first process is the inelastic parton scattering
mediated by the space-like gluons, known as Landau damping and 
the  second one is the gluon-dissociation process which
corresponds to the decay of a color singlet state into a
color octet induced by a thermal gluon~\cite{Brambilla:2013dpa}.
Recently two of us derived the imaginary component
of the potential using thermal field theory, where the inclusion
of confining string term makes the (magnitude) imaginary component 
larger, compared to the
medium-modification of the perturbative term
alone~\cite{Vineet:PRC2009,Lata:PRD2013,Lata:PRD2014}
However the results referred above about the imaginary component 
of the potential are limited to the weak coupling techniques. Using 
holographic 
correspondence, authors in \cite{Yavier:PRD2008,Hatsuda:PRD2013} in strong coupling limit
have obtained the imaginary part of the potential for
${\cal N} = 4$ supersymmetric gauge theory beyond a critical separation,
$r_c$ of a static $Q \bar Q$ pair,
by analytically continuing the string configurations into the
complex plane. They suggested that the complex-valued saddle points 
beyond $r_c$ (=$0.87 z_h$) may be interpreted as the quasi-classical 
configurations in the classically forbidden region of string coordinates, 
analogous to the method of complex trajectories
used in quasi-classical approximations to quantum mechanics \cite{LL3}.
Complex valued string coordinates at the extrema have also been considered in 
the calculation of  total cross section for deep inelastic scattering (DIS) in 
\cite{Yuri:JHEP2008}. In an alternative approach in AdS/CFT framework, the imaginary part 
is calculated in a strongly coupled plasma through the thermal 
worldsheet fluctuation method~\cite{Noronha:2013JHEP}. In the 
aforesaid calculations, the dual geometry taken was the pure AdS black hole
metric, hence their dual gauge theory is conformally invariant
unlike QCD which depends on scale.

A resurrected interest in the properties of bound states 
moving in a thermal medium arose in recent years due to
the advent of high energy heavy-ion colliders. 
Using holographic correspondence, the potential between two heavy 
quarks moving through the medium or, equivalently, two heavy quarks at 
rest in a moving medium is calculated in a pure AdS black hole 
background~\cite{Liu:2006nn, Chernicoff:2006hi,Avramis:2006em,
caceres:2006,Liu:2013}, where the solution above a critical separation 
of the pair is abandoned 
and is discussed in details in~\cite{Rajagopal}.
Recently the velocity dependence of the imaginary potential of 
the traveling bound states is also calculated~\cite{aliakbari:PRD2014},
following the idea of thermal worldsheet fluctuation 
method~\cite{Noronha:2013JHEP,fadafan:JHEP2013}. However in abovementioned 
calculations the metric used is AdS black hole, thus the dual gauge theory
is conformal, unlike QCD.
 	
Our aim is to improve the abovementioned calculations in two folds: first 
of all we improve the dual gravity which is somewhat closer to QCD than pure
AdS geometry in the sense that it accounts for RG flow
\cite{Mia:PRD2010}.
Secondly, unlike the world-sheet thermal fluctuations, we delve into 
the solution beyond the critical separation of the pair, which leads
to the complex-valued string configuartions and hence the potential 
turns out to be complex. Recently we have 
implemented both points in a calculation for a static pair~\cite{Binoy:PRD15},
and in the present work, we wish to extend the calculation
for a moving pair. Our paper is organised as follows : 
In Section 2, we have revisited the OKS-BH
geometry in brief and in Section 3, we employ this geometry to
obtain both the real and imaginary parts of the potential by the Nambu-Goto 
action. Thereafter we calculate the thermal width for the ground and 
excited states in Section 4. Finally we conclude in Section 5.

\section{Construction of dual geometry}
In this section we discuss briefly about the OKS-BH geometry that we will 
use to 
calculate the potential. The discussion presented here closely follows 
\cite{Mia:NPB2010,Mia:PRD2010,Mia:PLB2011} and the readers already aware 
of this 
geometry may go directly to next section to see the calculations.

A gauge theory is said to be conformal if it does not flow with the 
scale, {\em i.e.} has a trivial renormalisation group (RG) flow, and 
a conformal theory in a (3 + 1)-dimension can be conveniently 
described on the boundary of  pure anti-de Sitter 
space~\cite{Maldacena:1998adth}. But if the theory has 
a non-trivial RG flow like QCD, which is 
confining in IR while conformal in UV, we 
cannot describe the full theory on the boundary of some 
five dimensional space and hence need to look differently at 
running energy scales. One way out is to embed the D branes in the 
geometry and as a result
the corresponding gauge theory exhibit logarithmic RG flow. 
Such a construction was done in the Klebanov-Strassler (KS) warped 
conifold geometry \cite{KS}. The gravity dual for KS model is  
warped deformed conifold with three-form 
type IIB fluxes and the corresponding gauge theory is confining 
in the far IR but is not free at UV.
Other problems with KS model include the absence of quarks in the fundamental 
representation and the non applicability of model at non-zero temperature. 
We look at these one by one.

In stringy model of gauge dynamics, inclusion of fundamental matter is 
possible by embedding a set of flavor branes in  addition to the color 
branes. The strings connecting 
to the color branes in the adjoint representation of U($N_c$) group gives
gauge particles while those connecting to the flavor 
branes” in the adjoint representation of U($N_f$) gives
the mesons, and those connected to both the flavor 
and color  branes in the fundamental representation gives the quarks 
and anti-quarks, having the strings with the opposite orientation.

Theoretically one could go for both large $N_c$ and $N_f$ in near 
horizon limit and translates
the branes into fluxes and derive the gravity background that is a
holographic dual of a gauge system with gluons and quarks.
In practice one does it using probe approximation, where the backreaction 
of the probes on the background can be neglected
for the case $ N_f \ll N_c$ and the flavor physics is then extracted 
by analyzing the effective action, namely DBI action plus CS action 
that describes the flavor branes in the color background. This procedure 
was already adopted in ${AdS}_5 \times S^5$ model
\cite{Karch:2002}, in confining model~\cite{sakai:2003} etc.

To  introduce the fundamental quarks in the original Klebanov-Strassler 
model by embedding D7 branes in the gravity dual \cite{KS,Kuperstein:2005,Kuperstein:2009} is subtle
because the full global solution that incorporate back reaction of the 
D7 branes on the KS background is complicated and so far we only know how to insert coincident D7 branes in the 
Klebanov-Tseytlin background \cite{KS}. The resulting background 
is the Ouyang  background discussed in \cite{KS} which has
all the type IIB fluxes switched on including the axio-dilaton. In 
this OKS (Ouyang-Klebanov-Strassler) model,
the local metric was computed by incorporating the deformations 
of the seven branes
when these branes are moved far away from the regime of interest.
Since the seven branes are kept far away, the axion-dilaton
vanish for the background locally. However there will be
non-zero axion-dilaton globally. The local back reactions
on the metric would modify the warp factors to the full global scenario.

Now to switch on a non-zero temperature we need to insert a black hole into 
this background and the Hawking temperature will 
correspond to the gauge theory temperature.
Combining all physics ingredients together 
the metric in OKS-BH geometry looks like 
\begin{equation}
ds^2 = {1\over \sqrt{h}}
\Big[-g_1(u)dt^2+dx^2+dy^2+dz^2\Big] +\sqrt{h}\Big[g_2^{-1}(u) du^2+ d{\cal M}_5^2\Big]
\end{equation}
where $h$ is the warp factor, $g_i(u)$ are the black-hole 
factors, $u$ denotes the extra dimension and 
$d{\cal M}_5^2$ is the metric of warped resolved-deformed 
conifold. 
 
The above  picture works well at IR. In fact OKS model has emerged as the most promising candidate 
to study large N strongly coupled QCD, but the effective number of
degrees of freedom grow indefinitely in the UV. Moreover in the presence of 
fundamental flavors such a proliferation leads to 
Landau poles and hence the UV divergences of the Wilson loops. Our aim here 
is to detemine the Nambu-Goto action with 
a dual gravity of a gauge theory at finite temperature, which resembles
the main features of strongly coupled QCD, i.e., a gauge theory which 
is almost conformal in the UV with
no Landau poles or UV divergences of the Wilson loops, but has 
logarithmic running of coupling in the IR. In other words we want  
AdS-Schwarzschild geometry in asymptotic limit and therefore we need to add
the appropriate UV cap to it.

Now the addition of UV caps in general can change IR geometries. However 
the far IR geometries have not been changed
by the addition of UV caps because the UV caps correspond to adding
non-trivial irrelevant operators in the dual gauge theory. These operators 
keep far IR physics completely unchanged, but physics at not-so-small 
energies may change a bit. So the IR geometry part 
has been modified to obtain the desired dual gauge theory in the 
following steps~\cite{Mia:PRD2010,Mia:NPB2010,Mia:PLB2011}:\\
i) switch off the three-form fluxes at large $r$ so as to keep only 
five-form fluxes to get the AdS type solutions.\\
ii) modulate the axio-dilaton
behaviour so that the axio-dilaton field decay rapidly at large $r$ 
instead of the logarithmic growth, to get rid off the Landau poles.\\ 
iii) implement both of them by switching on anti D5-branes and
anti seven-branes with electric and magnetic fluxes on them to 
kill off the unnecessary
tachyons and thereby restore the stability. Additionally the anti 
seven-branes, along with
the seven-branes, were embedded in a non-trivial way so as not 
to spoil the small $r$, or
the IR, behaviour of the theory. The anti D5-branes are dissolved 
on the seven-branes, but they contribute to the three-form fluxes. 
Both of them together control both the three form
fluxes as well as the axio-dilaton at large $r$.

As a consequence, the above metric will receive corrections as
\begin{equation} \label{eq:correction}
ds^2 = {1\over \sqrt{h}}
\Big[-g(u)dt^2+dx^2+dy^2+dz^2\Big] +\sqrt{h}\Big[g(u)^{-1}g_{uu}du^2+ g_{mn}dx^m dx^n\Big]
\end{equation}
where we have set black hole factors $ g_1(u) = g_2(u) = g(u)$. The corrections $ g_{uu} $ 
are of the form $ u ^{-n} $ and appear in the metric because the 
existence of axio-dilaton and the seven-brane sources tell us that the unwarped metric 
may not remain Ricci flat. Thus the corrections may be written as  
\begin{equation}
g_{uu} = 1 + \sum_{i = 0}^{\infty} \frac{a_{uu,i}}{u^i}
\end{equation}
where $ a_{uu,i}$ are the coefficients independent of the coordinate u and 
can be solved for exactly as shown in \cite{Mia:PRD2010}.
The warp factor, $h$ can be obtained as
\begin{equation*}
 h ~= ~\frac{L^4}{u^4}\left[1+\sum_{i=1}^\infty\frac{a_i}{u^i}\right]~~~ 
\end{equation*}
where $a_i $ are coefficients, again independent of the coordinate u. These  
are of ${\cal O}(g_sN_f)$ and can be solved 
for exactly as shown in \cite{Mia:PRD2010} and $L$
 denotes the curvature of space. 
This metric reduces to OKS-BH in IR and is asymptotically $AdS_5$ x 
$M_5$ in the UV. It describes the geometry all the way from the IR to the
UV.

With the change of coordinates $ z = 1/u $, we can rewrite the metric
(\ref{eq:correction}) as
\begin{equation}\label{eq: metric}
ds^2 ~=~ g_{\mu\nu} dX^\mu dX^\nu ~=~
 A_n z^{n-2}\left[-g(z)dt^2+d\overrightarrow{x}^2\right]
~ +~\frac{
B_l z^{l}}{
 A_m z^{m+2}g(z)}dz^2+\frac{1}
{A_n z^{n}}~ds^2_{{M}_5},
\end{equation}
where $ds^2_{{\cal M}_5}$ is the metric of the internal space 
and $A_n$'s are the coefficients that can be extracted from the 
$a_i$'s as follows: 
\begin{equation}
{1\over \sqrt{h}}~ = ~ {1\over L^2 z^2 \sqrt{a_i z^i}} \equiv  
A_n z^{n-2} ~=~ {1\over L^2 z^2}\left[a_0  
-{a_1z\over 2} + \left({3a_1^2\over 8a_0} - {a_2\over 2}\right)z^2 
+ \cdot \cdot \cdot \right] \quad,
\end{equation}
which gives $ A _0 = {a_0\over L^2},  A_1 = -{a_1\over 2L^2},  
A_2 = {1\over L^2}\left({3a_1^2\over 8a_0} - {a_2\over 
2}\right)$ and so on. Note that since $a_i$'s for  $i \ge 1$ 
are of ${\cal O}(g_sN_f)$ and $L^2 \propto \sqrt{g_sN}$, so 
in the limit $ g_s N_f 
\rightarrow 0$ and $ N \rightarrow \infty $ all $ 
A_i$'s for $ i \ge 1$ are very small. The $u^{-n}$ corrections mentioned 
above in Eq.(\ref{eq:correction}) are accommodated via $ B_l z^l$ series which is given by  
\begin{equation}
B_l z^l = 1 + a_{zz,i} z^i
\end{equation}

In fact the complete picture can be divided into three regions. Region 1 is 
the IR region where we have pure OKS-BH geometry. Region 3 is the UV 
region where UV cap has been added. And region 2 is the interpolating 
region between UV and IR. 
The background for all these three regions and the process of adding 
UV cap has been described in full details in \cite{Mia:PRD2010}. Also the RG 
flow associated with these regions and the corresponding field theory 
realizations have been discussed in \cite{Chen:PRD2013}.
We shall not go into 
the complete details here and  will use the metric given
in Eq.(\ref{eq: metric}) in extremizing the action 
to calculate the potential in the next section. 

Another model that is extensively used to study certain IR 
dynamics of QCD is the Sakai-Sugimoto model\cite{sakai:model} which
is a beautiful brane construction in type IIA theory.
The model consists of a set of N wrapped color D4-branes on
the circle, while the flavor branes D8 and $\bar{D8}$ 
placed at the anti-nodal points of the circle give
the mesonic states. In the gravity dual, the wrapped D4-branes 
are replaced by a geometry, i.e an asymptotically
AdS space, but the eight-branes remain and so does the circular 
direction. However the Sakai-Sugimoto model does not have a UV 
completion, which is very important for the study of heavy 
quarkonium. The UV contribution to the the heavy quark potential
for the heavy quark bound states are dominant compared to
IR part because since their mass is larger than QCD scalei so the
relative velocity is very small and becomes weakly coupled
($v \sim \alpha_s$, $\alpha_s$ is the strong interaction
coupling). The comparison of Sakai-Sugimoto model with the one 
that we use here has recently been done in \cite{Mia:2015}. 

\section{Potential between a heavy quark and a heavy antiquark}
The potential at finite temperature can 
be obtained from the Wilson loop's expectation value 
evaluated in a thermal state of the gauge theory :
\begin{equation}
\label{eq:wilsonrec1}
\lim_{\mathcal{T} \to \infty}\langle W(C) \rangle 
\sim e^{i \mathcal{T} V_{Q\bar{Q}}(r,T)}
\end{equation}
According to the gauge/gravity prescription \cite{Maldacena:1998adth}, the 
expectation value of $W(C)$ in a strongly coupled gauge theory 
dual to a theory of gravity is
\begin{equation}
\label{eq:wilsongaugegravity}
\langle W(C) \rangle \sim Z_{str} \sim e^{i S_{str}}
\end{equation}
where $S_{str}$ is the classical string action propagating 
in the bulk evaluated at an extremum. Hence the potential can be obtained 
by extremizing the world-sheet 
of open string attached to the heavy quark pair located at the 
boundary of ${\rm{AdS}}_5$ space in the background of ${\rm{AdS}}_5$ space
and in the background of ${\rm{AdS}}_5$ 
black-hole metric, for determining the potentials at $T=0$ and $T\ne 0$,
respectively.

We consider a quark-antiquark pair ($Q \bar Q$) moving along $x_3$ 
direction with some 
velocity $\vec{v}$. It is convenient to boost the system to the rest frame ($t',
x_3'$) of the quark-antiquark pair as
\begin{eqnarray}\label{boost} 
       dt &=& dt^\prime \cosh ~ \eta - dx_3' \sinh ~ \eta,  \nonumber \\
       dx_3 &=& -dt^\prime \sinh~ \eta + dx_3' \cosh ~ \eta,
\end{eqnarray}
with $ \tanh~\eta = v$. The $ Q \bar{Q} $ dipole is now at rest but the quark-
gluon plasma is moving with velocity $\vec{v}$ in the negative $x_3'$- 
direction. Thus under the boost, the metric (\ref{eq: metric}) reduces to
\begin{eqnarray}\label{eq: metricboost}
ds^2  &=&
 A_n z^{n-2}\bigg[dt^2\left(-1 + \frac{z^4}{z_h^4} \cosh^2 \eta \right) + dx_1^2 + dx_2^2 
 + 
 dx_3^2\left(1 + \frac{z^4}{z_h^4} \sinh^2 \eta \right) \nonumber \\ 
&-& 2~dt~dx_3~\sinh \eta~\cosh \eta~ 
 \frac{z^4}{z_h^4}  \bigg] + \frac{
B_l z^{l}}{
 A_m z^{m+2}g(z)}dz^2+\frac{1}
{A_n z^{n}}~ds^2_{{M}_5}
\end{eqnarray}
Now the Nambu-Goto action can be defined in terms of string coordinates
($\sigma$, $\tau$):
\begin{equation}
S_{NG} = \frac{1}{2 \pi} \int d \sigma d \tau \sqrt{-
det \left[(g_{\mu \nu} + \partial_\mu \phi \partial_\nu 
\phi)\partial_a X^{\mu} \partial_b X^{\nu}\right]}~,
\end{equation}
where $\phi$ is the background dilaton field, which is responsible 
for breaking of conformal symmetry of theory and is given by 
\begin{equation}
\phi = \log g_s - g_s D_{n + m_o} z^{n + m_o}
\end{equation}
We parametrize the two-dimensional world sheet as
\begin{equation}
~~~ \tau = t, ~~~\sigma = x_1 \in [-\frac{r}{2}, \frac{r}{2}], ~~~ x_2 = 
{\rm{const}}, ~~~ x_3 = 
{\rm{const}}, ~~~ z = z(x_1), ~~~ \partial_a = \frac{\partial}{\partial 
\tau}, ~~~ \partial_b = \frac{\partial}{\partial \sigma}~,
\end{equation} 
thus the Nambu-Goto action can be rewritten as   
\begin{equation}
S_{NG} = \frac{\mathcal{T}}{2 \pi} \int_{-\frac{r}{2}}^{\frac{r}{2}} dx_1 
\sqrt{(A_n z^n)^2 \left(\frac{1}{z^4}-\frac{1}{z_h^4} \cosh^2 \eta \right) 
+ B_m z^m \frac{(z')^2}{z^4} ~ 
\frac{(z_h^4 - z^4 \cosh^2 \eta ) }{(z_h^4- z^4)}} ~,
\end{equation}
where we have used
\begin{equation*}
B_mz^m = B_l z^l + 2 g(z)g_s^2(n + m_o)D_{n + m_o}(l + 
m_o)D_{l + m_o} A_k z^{k + n+ l + 2m_o}~.
\end{equation*} 

The Nambu-Goto action can also be written as an integral over $z$,
\begin{equation} \label{eq: action}
S_{NG} = \frac{\mathcal{T}}{\pi} \int_0^{z_{max}} \frac{d z}{z'} 
\sqrt{(A_n z^n)^2 \left(\frac{1}{z^4}-\frac{1}{z_h^4} \cosh^2 \eta \right) 
+ B_m z^m \frac{(z')^2}{z^4} ~ \frac{(z_h^4 - z^4 \cosh^2 \eta ) }{(z_h^4- z^4)} } 
\end{equation}
Since this action does not depend explicitly on $x$, so the corresponding 
Hamiltonian will be a constant of motion, i.e., 
\begin{equation}
H = z' \frac{\partial L}{\partial z'} - L = C_0~({\rm{say}})~,
\end{equation} 
where the constant $C_0$ can be determined from the following equation: 
\begin{equation*}
\frac{1}{L}\left( (z')^2 \frac{B_m z^m}{z^4}~\frac{(z_h^4 - z^4 \cosh^2 \eta ) }{(z_h^4- 
z^4)}   -L^2\right) = - \left(\frac{1}
{z^4} - \frac{1}{z_h^4} \cosh^2 \eta \right) \frac{(A_n z^n)^2}{L} = C_0 
\end{equation*}

Since at $ z = z_{\rm{max}}$, $z^\prime$=0, where $z_{\rm{max}}$ is the 
maximum of the string coordinate along the fifth dimension, so we can 
find out the derivative $z^\prime$ by the simple algebra
\begin{eqnarray} \label{eq: z der}
z' = \frac{d z}{d x} &=& \frac{(A_n z^n)^2 ~ z_{max}^2}{
\sqrt{B_m z^m} ~ z_h^2 ~ z^2 (A_n z_{max}^n)} \sqrt{\frac{(z_h^4 -z^4)(z_h^4 -z^4 \cosh^2 
\eta) 
}{(z_h^4 -z_{max}^4 ~ \cosh^2 \eta)}} \nonumber \\ 
 & \times &  \left( 1 - \frac{(A_n z_{max}^n)^2 (z_h^4 -z_{max}^4~ 
\cosh^2 \eta) z^4 }{(A_n z^n)^2 (z_h^4 - z^4 \cosh^2 \eta) z_{max}^4} \right)^\frac{1}{2} 
\end{eqnarray}
Integrating both sides, the above equation yields the separation of the
$Q \bar Q$ pair, $r$ as 
\begin{eqnarray}
r &=& \frac{2 z_h^2 \sqrt{z_h^4 -z_{\rm{max}}^4~ \cosh^2 \eta} ~ (A_n 
z^n_{\rm{max}})}
{z_{\rm{max}}^2}  \int_0^{z_{\rm{max}}} dz \frac{z^2 \sqrt{B_m z^m}}{\sqrt{(z_h^4 
- z^4)(z_h^4 - z^4 \cosh^2 \eta)} ~ (A_n z^n)^2}  \nonumber \\
& \times & \left(1 - \frac{(z_h^4 -z_{max}^4 \cosh^2 \eta )(A_n z_{max}^n)^2 z^4}
{(z_h^4 -z^4 \cosh^2 \eta ) ~ (A_n z^n)^2 ~ z_{max}^4} \right)^{-\frac{1}{2}} 
\end{eqnarray}
Expanding the square root and keeping only the first term  because 
the coefficients $A_i$'s are small, the separation ($r$) becomes
\begin{equation} \label{eq: r-z}
r = \frac{2 z_h^2 (A_n z^n_{max})~\sqrt{z_h^4 -z_{max}^4 \cosh^2 \eta}~}
{z_{max}^2} I,
\end{equation}
where the integral $I$ is defined by
\begin{equation} \label{eq:integral}
I = \int_0^{z_{max}}dz \frac{z^2 \sqrt{B_m z^m}}{(A_n z^n)^2 \sqrt{(z_h^4 - 
z^4)(z_h^4 - z^4 \cosh^2 \eta)} }
\end{equation}

\par
Now we obtain the action 
\begin{equation*}
S_{NG} = \frac{\mathcal{T}}{\pi} \int_0^{z_{max}} \frac{d z }{z^2}\sqrt{\frac{B_m 
z^m(z_h^4-z^4 \cosh^2 \eta)}{(z_h^4 - z^4)}}\left(1 - \frac{(z_h^4 -z_{max}^4~ 
\cosh^2 \eta 
)(A_n z_{max}^n)^2 
z^4}{(z_h^4 -z^4 \cosh^2 \eta )(A_n z^n)^2 z_{max}^4} \right)^{-\frac{1}{2}} 
\end{equation*}
Expanding the square root and keeping only the first two terms (since the 
coefficients $A_i$'s are small), the above action can be written as
\begin{equation}
S_{NG} = \frac{\mathcal{T}}{\pi} \left[ \int_0^{z_{max}} \frac{d z 
}{z^2} \sqrt{\frac{B_m 
z^m(z_h^4-z^4 \cosh^2 \eta)}{(z_h^4 - z^4)}} ~ + ~ \frac{1}{2}\frac{(z_h^4 -z_{max}^4 ~ 
\cosh^2 \eta)(A_n 
z_{max}^n)^2}{z^4_{\rm{max}}}~ I \frac{}{} \right]~,
\end{equation}
where $I$ is the same integral as in (\ref{eq:integral}). Thus substituting 
the integral $I$ in terms of $r$, the action becomes
\begin{eqnarray}
S_{NG}  &=&  \frac{\mathcal{T}}{\pi} \int_0^{z_{max}} \frac{d z}{z^2}\sqrt{\frac{B_m 
z^m(z_h^4-z^4 \cosh^2 \eta)}{(z_h^4 - z^4)}} ~+~ \frac{\mathcal{T}}{4 \pi} 
\frac{\sqrt{z_h^4 -z_{max}^4 ~ \cosh^2 \eta }}
{z_h^2 ~ z^2_{max}} A_n z^n_{max} ~ r\nonumber\\
&\equiv & S^{1} + S^{2} 
\end{eqnarray}
The first term in the action, $S^1$ diverges in the lower limit of the
integration, so we regularize it by integrating from $\epsilon$ (instead
of 0) to $z_{\rm{max}}$~\cite{Mia:PRD2010} and then identifying the divergent 
term in the integral
\begin{equation*}
S^{1} = \frac{\mathcal{T}}{\pi} \int_{\epsilon}^{z_{max}} 
\frac{d z \sqrt{B_m z^m}}{z^2} \sqrt{\frac{z_h^4 - z^4 \cosh^2 \eta}{z_h^4 - z^4}} \quad ,
\end{equation*}
Let us also assume for simplicity that 
the higher coefficients $A_i $ and $B_i $ are very small for $i \geq 
3$ and can be neglected. This simplification reduces the series 
$ A_n z^n$ to $ 1 + A_2 z^2 $ and the other series $ B_m z^m $ to 
$ 1 + B_2 z^2 $, where we take $A_0 = 1$ and $A_1 = 0$ and similarly 
$ B_0 = 1$ and $B_1 = 0 $. This will also simplify all 
the expressions and helps us to keep an analytic control on the 
equations. Therefore, in the limit of small coefficients, $S^1$ becomes
\begin{equation}
S^{1} = \frac{\mathcal{T}}{\pi} \int_{\epsilon}^{z_{max}} \frac{d z (1 + \frac{B_2}{2} 
z^2)}{z^2} \sqrt{\frac{z_h^4 - z^4 \cosh^2 \eta}{z_h^4 - z^4}} \equiv
\frac{\mathcal{T}}{\pi} \int_{\epsilon}^{z_{max}} d z~ F(z,v)~,
\end{equation}
where the integral, $F(z,v)$ can be expanded in the 
Taylor series in velocity:
\begin{equation*}
F(z,v) = F(z,0) + \frac{\partial F}{\partial v} v +  \frac{1}{2}
\frac{\partial ^2 F}{\partial v^2}~{v^2} + \cdot \cdot \cdot 
\cdot \cdot  \cdot \cdot \cdot
\end{equation*}

We can omit the dotted terms in the small velocity limit and can write $S^{1}$ 
as 
\begin{equation}
S^{1} = \frac{\mathcal{T}}{\pi} \left[ \int_{\epsilon}^{z_{max}} d z \left( \frac{1}{z^2} 
+ \frac{B_2}{2} \right) -\frac{v^2}{2} \int_{\epsilon}^{z_{max}} d z \left( z^2 + 
\frac{B_2}{2} z^4 
\right) \frac{1}{(z_h^4 - z^4)} \right]
\end{equation}
After subtracting the divergent piece in the limit
$\epsilon \rightarrow 0$, we get the renormalised Nambu-Goto action 
\begin{eqnarray}
S^{\rm{ren}}_{\rm{NG}} &=& \frac{\mathcal{T}}{\pi} \bigg[\left(\frac{B_2}{2} z_{max} - \frac{1}
{z_{max}} 
\right) - \frac{v^2}{16 z_h} \bigg( -4 B_2 z_h z_{max} ~+~ 2 (-2 + B_2 z_h^2 ) \tan^{-1} 
\left(\frac{z_{max}}{z_h}\right)  \nonumber \\  &+&
 (2 + B_2 z_h^2 ) \log \frac{(z_h + z_{max})}{(z_h - z{max})} \bigg) 
\bigg]~+~ \frac{\mathcal{T}}{4 \pi} \frac{\sqrt{z_h^4 -z_{max}^4 ~ \cosh^2 \eta }}
{z_h^2 ~ z^2_{max}} A_n z^n_{max} ~ r
\end{eqnarray}
and hence the potential is obtained by
\begin{eqnarray}\label{eq:pot}
V_{Q\bar{Q}} &=& \lim_{\mathcal{T} \rightarrow \infty} \frac{S^{\rm{ren}}_{\rm{NG}}}{\mathcal{T}} 
\nonumber \\
&=& \frac{1}{\pi} \bigg[\left(\frac{B_2}{2} z_{max} - \frac{1}{z_{max}} 
\right) - \frac{v^2}{16 z_h} \bigg( -4 B_2 z_h z_{max} ~+~ 2 (-2 + B_2 z_h^2 ) \tan^{-1} 
\left(\frac{z_{max}}{z_h}\right)  \nonumber \\  &+&
 (2 + B_2 z_h^2 ) \log \frac{(z_h + z_{max})}{(z_h - z{max})} \bigg) 
\bigg]~+~ \frac{1}{4 \pi} \frac{\sqrt{z_h^4 -z_{max}^4 ~ \cosh^2 \eta }}
{z_h^2 ~ z^2_{max}} A_n z^n_{max} ~ r~.
\end{eqnarray}

The potential thus obtained are functions of both $z_{\rm{max}}$ and $r$,
which are redundant. To make the potential as a function of $r$ only, we 
will express $z_{\rm{max}}$ as a function of $r$ 
and then plug in to the above potential. To do that we first concentrate 
on the integral $I$ (\ref{eq:integral}) which, keeping upto 
the second-order in both series in $z$, reduces to:
\begin{eqnarray} \label{eq: I}
I &=& \int_0^{z_{max}}dz \frac{z^2 \sqrt{1 + B_2 z^2}}{\sqrt{(z_h^4 - 
z^4)(z_h^4 - 
z^4 \cosh^2 \eta)}~ (1 + A_2 z^2)^2} \nonumber\\
& \equiv& \int_{0}^{z_{max}} d z ~ I'(z,v)
\end{eqnarray} 
In the limit of small velocity, we expand the integrand $I'(z,v)$ in 
the Taylor series in velocity, $v$ as: 
\begin{equation*}
I'(z,v) = I'(z,0) + \frac{\partial I'}{\partial v} v +  \frac{1}{2}
\frac{\partial ^2 I'}{\partial v^2}~ {v^2} + \cdot \cdot 
\end{equation*} 
and neglecting the higher order terms, the integral can be written as:
\begin{equation*}
I = \int_0^{z_{max}}dz \frac{z^2 \sqrt{1 + B_2 z^2}}{{(z_h^4 - 
z^4)} (1 + A_2 z^2)^2} + \frac{v^2}{2} 
\int_0^{z_{max}}dz \frac{z^6 \sqrt{1 + B_2 z^2}}{{(z_h^4 - 
z^4)^2} (1 + A_2 z^2)^2}
\end{equation*}

We will now find the solution for the separation, $ r$ in two limits, 
{\em namely} $z_{max} >> z_h $ and $z_{max} << z_h $. First for the 
$z_{max} >> z_h $ limit, the separation $r$ from Eq.(\ref{eq: r-z}) 
becomes as a function of $z_{\rm{max}}$
\begin{eqnarray}
\label{rlargezmax}
\frac{r \sqrt{(1-v^2)}}{2 i z_h^2} &=& -(c_1 + i c_2 - \frac{v^2}{2} (c_3 + i c_4 ))( A_2  
z^2_{max} 
+ 1) + \frac{B_2 (1-\frac{v^2}{2}) }{6 A_2 ~ z_{max}}  \nonumber \\ &+&  \frac{1}
{2}\frac{z_h^4~ A_2 
(1-v^2)}{z^2_{max}}
 (c_1 + i c_2 - \frac{v^2}{2} (c_3 + i c_4 ) ) +  \frac{(6 
A_2 - B_2)(1-\frac{v^2}{2})}{30 ~A_2^2 ~z^3_{max}} \nonumber \\ &+& \frac{(1-v^2)}
{2}\frac{z_h^4 }
{z_{max}^4} (c_1 + i c_2 - \frac{v^2}{2} (c_3 + i c_4 )) 
\end{eqnarray}
where $c_1$ , $c_2$ , $c_3$ and $c_4$ are defined as 
\begin{eqnarray} \label{eq: c}
c_1 &=&  \frac{\pi}{8} \left[\frac{1}{z_h}\frac{(2-
B_2 z_h^2)}{(A_2 z_h^2 - 1)^2}  - \frac{(6 A_2 - B_2 + A_2^2 z_h^4(2 A_2 
-3 B_2))}{\sqrt{A_2} (-1 + A_2^2 z_h^4)^2 } \right] \\  
 c_2 &=&  \frac{\pi}{8 z_h} \frac{(2 + B_2 z_h^2)}{(A_2 
z_h^2 + 1)^2}  \\ 
c_3 &=& \frac{\pi}{8} \left[ \frac{(6 A_2 - B_2 + A_2^2 z_h^4(10 A_2 
-7 B_2))}{\sqrt{A_2} (-1 + A_2^2 z_h^4)^3 } - \frac{(6 - 5 B_2 z_h^2 + A_2 z_h^2 (2 + B_2 
z_h^2))}{4 z_h (A_2 z_h^2 - 1)^3 } \right] \\
c_4 &=& \frac{ \pi (6 + 5 B_2 z_h^2 + A_2 z_h^2 (-2 + B_2 z_h^2))}{32 z_h (A_2 z_h^2 + 
1)^3 }
\end{eqnarray} 
We now invert the series (\ref{rlargezmax}) to obtain $ z_{max} $ in terms of 
$r$ as, 
\begin{eqnarray} \label{eq: zmax-r1}
z_{max} &=& \frac{1}{\sqrt{2 A_2 z_h^2(-i c_1 + c_2 + \frac{v^2}{2}(i c_3 
- c_4) )}} \sqrt{r \sqrt{(1 - v^2)}} \nonumber \\
&-& \sqrt{ \frac{(-i c_1 + c_2 + \frac{v^2}{2}(i c_3 - c_4)) z_h^2}{2 A_2}} 
\frac{1}{\sqrt{r \sqrt{(1 - v^2)}}} 
~-~ \frac{i}{12}\frac{B_2}{A_2} \frac{z_h^2 (2-v^2) }{r\sqrt{(1 - v^2)}}~,
\end{eqnarray}
which shows that the string coordinates become complex\footnote{
This solution was earlier abandoned in evaluating the potential for
the heavy quark pair in a moving medium~\cite{Rajagopal}.} and hence the potential from Eq. 
(\ref{eq:pot})
becomes imaginary:
\begin{eqnarray}
V_{Q \bar{Q}} &=& \frac{1}{\pi} \left( \frac{B_2}{2} z_{max}(1+\frac{v^2}{2}) - \frac{\pi 
v^2}{16 
z_h} (-2 + B_2 z_h^2 -i(2 + B_2 z_h^2))  - \frac{1}{z_{max}}(1+\frac{v^2}{2}) -\frac{v^2}
{12} 
\frac{z_h^4 B_2}{z_{max}^3}  \right) \nonumber \\ &-& \frac{i r}{4 \pi z_h^2 \sqrt{(1 - 
v^2)}} \left(A_2 z^2_{max} + 1 - \frac{A_2}{2} \frac{z_h^4 ~ (1 - v^2)}{z^2_{max}} - 
\frac{1}{2} \frac{z_h^4 (1 - v^2)}{z_{max}^4} \right)   
\end{eqnarray}
After substituting $z_{max}$ in terms of $r$ from 
Eq.(\ref{eq: zmax-r1}),
\begin{figure}
\begin{center}$
\begin{array}{cc}
\includegraphics[keepaspectratio,width=3.0in,height=3.0in]{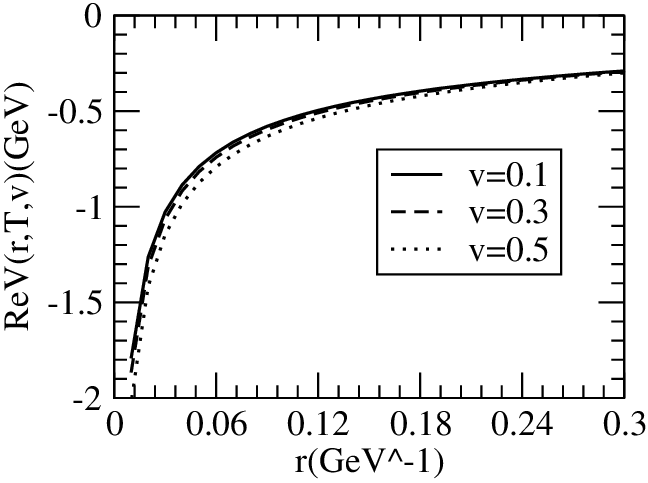} 
\hspace{0.25in}
\includegraphics[keepaspectratio,width=3.0in,height=3.0in]{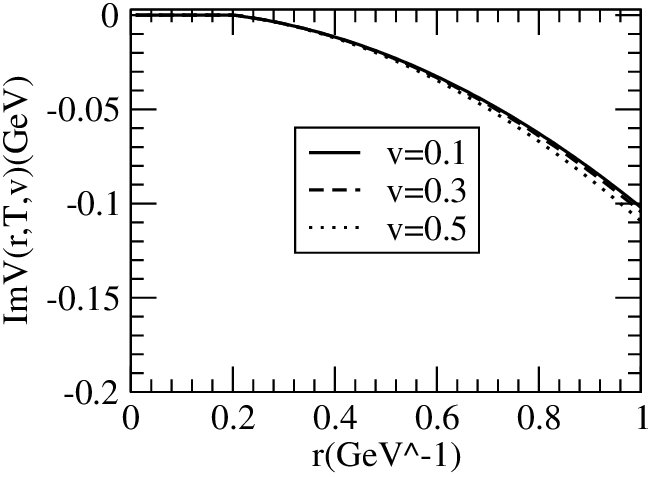} \\
\end{array}$
\end{center}
\caption{\footnotesize Variations of the real and imaginary part of 
the potential with the separation $r$ at various velocities (at a constant 
temperature 400 MeV), where both coefficients, $A_2$ and $B_2$ are taken as 0.24.}
\end{figure}
we obtain 
the real and imaginary parts of potential as
\begin{eqnarray}
Re[V(r,T,v)] &=&  \frac{r^2~(c_1 - \frac{v^2}{2} c_3)}{8 \pi z_h^4 (~(c_2- \frac{v^2}{2} 
c_4)^2 + (c_1 - \frac{v^2}{2} 
c_3)^2 ) } ~+~ \frac{(10-5 v^2 -6 v^4)}{24(1-v^2)} \nonumber \\ 
 & \times & \sqrt{\frac{\sqrt{(~(c_2- \frac{v^2}{2} c_4)^2 + (c_1 - \frac{v^2}{2} c_3)^2 
 )} + (c_2- \frac{v^2}{2} c_4) }
 {2 (~(c_2- \frac{v^2}{2} c_4)^2 + (c_1 - \frac{v^2}{2} c_3)^2 )}} \frac{B_2}{\sqrt{2 A_2} 
 \pi z_h } \sqrt{r 
 \sqrt{1-v^2}} \nonumber \\ 
  &-& \frac{v^2}{16 z_h} (-2 + B_2 z_h^2 ) ~-~ \frac{(4 A_2 + B_2 ) z_h}{\sqrt{2 A _2} \pi 
  } \frac{(18-9 v^2 -10 v^4)}{40 (1 - v^2 )} \nonumber \\
  & \times & \sqrt{\frac{\sqrt{(~(c_2- \frac{v^2}{2} c_4)^2 + (c_1 - \frac{v^2}{2} c_3)^2 
  )} + (c_2- \frac{v^2}{2} c_4) 
  }{2}} \frac{1}{\sqrt{r \sqrt{1- v^2}}}   
\end{eqnarray}
and 
 \begin{eqnarray}
Im[V(r,T,v)] & = & -\frac{r^2~(c_2 - \frac{v^2}{2} c_4)}{8 \pi z_h^4 (~(c_2- \frac{v^2}
 {2} c_4)^2 + (c_1 - 
 v^2 c_3)^2 ) } ~+~ \frac{(10-5 v^2 -6 v^4)}{24(1-v^2)} \nonumber \\ 
 & \times & \sqrt{\frac{\sqrt{(~(c_2- \frac{v^2}{2} c_4)^2 + (c_1 - \frac{v^2}{2} c_3)^2 
 )} - (c_2- \frac{v^2}{2} c_4) }
 {2 (~(c_2- \frac{v^2}{2} c_4)^2 + (c_1 - \frac{v^2}{2} c_3)^2 )}} \frac{B_2}{\sqrt{2 A_2} 
 \pi z_h } \sqrt{r 
 \sqrt{1-v^2}}~,
 \end{eqnarray}
respectively.
We have plotted the variation of the real and imaginary parts of 
potential with the separation $r$ for different velocities. We find that 
the potential becomes stronger with the increase of velocity, which
can be understood from the fact that the effective 
temperature gets reduced due to the recession of the medium and 
hence the screening becomes weaker. The imaginary part increases in magnitude as the velocity 
increases.
We will now find the potential for the 
other extreme limit {\em i.e.}, very small $ z_{max} $ 
and $z_{max} << z_h $. 
\begin{figure}
\begin{center}$
\begin{array}{cc}
\includegraphics[keepaspectratio,width=3.0in,height=3.0in]{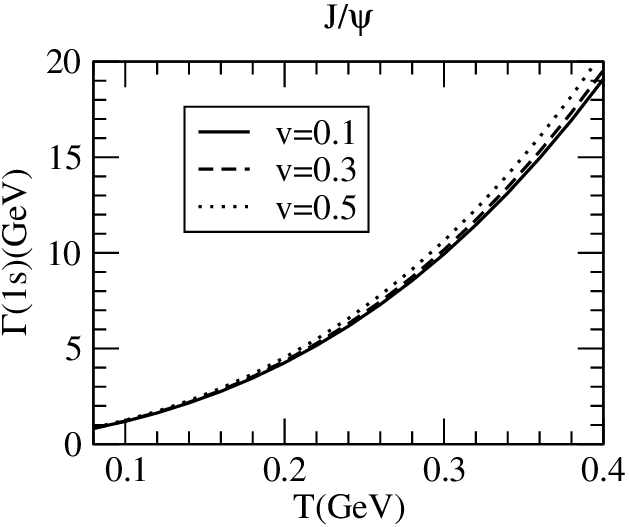} 
\hspace{1.5cm}
\includegraphics[keepaspectratio,width=3.0in,height=3.0in]{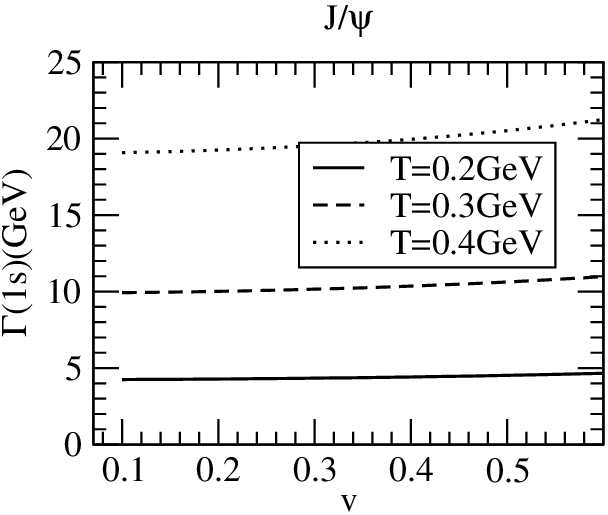}
\end{array}$
\end{center}
\caption{\footnotesize Variations of the thermal width for the charmonium 
ground state with the temperatures at increasing velocities (left) and 
with the velocity at increasing temperatures (right).}
\label{charm1s}
\end{figure}
In this limit, the separation $r$ becomes
\begin{equation}
r = \frac{2}{3} z_{\rm{max}} + \frac{(3 B_2 - 2 A_2)}{15} z_{\rm{max}}^3~,
\end{equation}
which will eventually be inverted to express $z_{max}$ in terms of 
$r$ as
\begin{equation} \label{eq: zmax-r}
\begin{split}
z_{max} = \frac{3 r}{2} ~+~ \frac{27}{80} ( 2 A_2 - 3 B_2) r^3 ~+~ O[r]^5 
\end{split}
\end{equation}
Thus the potential in this limit reduces to 
\begin{equation}
\begin{split}
Re [V_{Q \bar{Q}} (r,T)] \stackrel{z_{\rm{max}} \ll z_h}{\simeq}
-\frac{5}{9 \pi r} 
+ \frac{9 (A_2 + B_2)}{20 \pi} r + O[r]^3
\end{split}
\end{equation} 
Hence upto first order in $r$, the potential is independent of velocity 
in this limit.
The first term here is like the coulombic term while the 
second term is the linear confining term.
\section{Calculation of thermal width}
The thermal width $\Gamma_{Q\bar{Q}}$ arises due to the imaginary component
of the potential and is defined as
\begin{equation}
\label{eq:thermalwidth}
\Gamma_{Q\bar{Q}} = - \langle \psi | \mathrm{Im} \, V_{Q\bar{Q}}(r,T,v) | \psi \rangle,
\end{equation}
where the wave function for the ground state (1S) is given by
\begin{equation}
\label{eq:wavefunction}
\psi = \frac{1}{\sqrt{\pi} a_0^{3/2}} e^{-r/a_0}
\end{equation}
and for the first excited state
\begin{equation}
\label{eq:wavefunction1}
\psi = \frac{1}{4\sqrt{2\pi} a_0^{3/2}}\left( 2 - \frac{r}{a_0} \right) e^{-r/2 a_0}
\end{equation}
in a Coulomb-like potential, with $a_0 = {(\mu e^2)}^{-1}$ being the Bohr 
radius ($\mu$ is the reduced mass and $e^2$ is the square of the
electric charge). 
\begin{figure}
\begin{center}$
\begin{array}{cc}
\includegraphics[keepaspectratio,width=3.0in,height=3.0in]{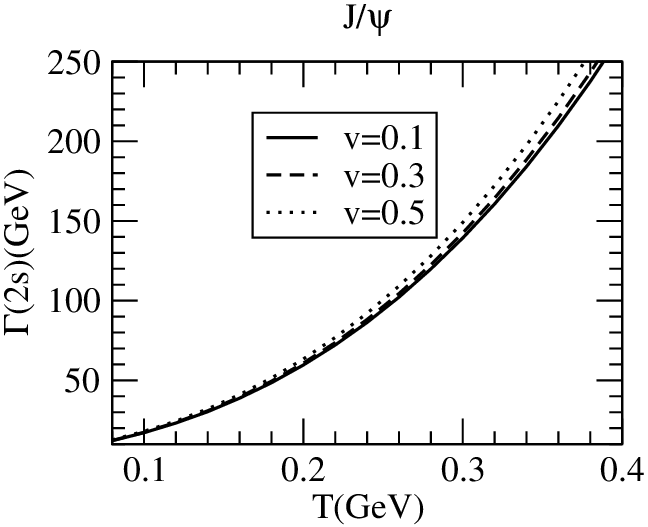} 
\hspace{1.1cm}
\includegraphics[keepaspectratio,width=3.0in,height=3.0in]{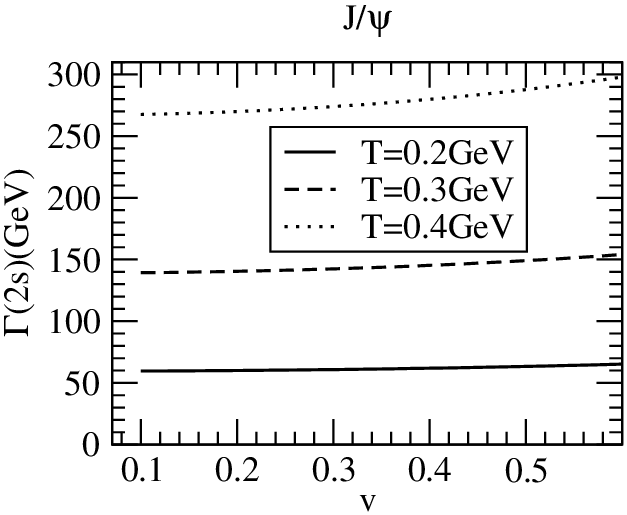}
\end{array}$
\end{center}
\caption{\footnotesize Variations of the thermal width for the charmonium 
first excited (2S) state with the temperatures at increasing velocities (left) and 
with the velocity at increasing temperatures (right).}
\label{charm2s}
\end{figure}
Even though the real part of the potential 
is not purely Coulombic but the leading contribution for the potential 
for the deeply bound $Q\bar{Q}$ states in a conformal plasma and 
thus justifies the use of Coulomb-like wave functions to determine the width.
Thus we calculate the thermal width for the ground-state(1S) as
\begin{eqnarray}
\label{eq:thermalwidth2}
\Gamma_{Q\bar{Q}} (1S) &=&-\frac{4}{a_0^3} \int_0^{\infty} dr \, r^2 e^{-2r/a_0}\, 
\mathrm{Im} \, 
V_{Q\bar{Q}}(r,T) \nonumber\\
&=& \frac{972 \pi^5 T^4}{200 m_Q^2} \frac{(c_2 - \frac{v^2}{2} c_4)}{((c_2- \frac{v^2}{2} 
c_4)^2 + (c_1 - \frac{v^2}{2} c_3)^2 )} ~-~ \frac{45 \sqrt{2} \pi T B_2 }{32 \sqrt{5 m_Q A_2}} {(1-
v^2)}^{\frac{1}{4}} \frac{(10-5 v^2 -6 v^4)}{24(1-v^2)} \nonumber \\ 
&\times& \sqrt{\frac{\sqrt{(~(c_2- \frac{v^2}{2} 
c_4)^2 + (c_1 - \frac{v^2}{2} c_3)^2 )} - (c_2- \frac{v^2}{2} c_4) }
 {2 (~(c_2- \frac{v^2}{2} c_4)^2 + (c_1 - \frac{v^2}{2} c_3)^2 )}}
\end{eqnarray}
and for the first excited state (2S), 
\begin{eqnarray}
\Gamma_{Q\bar{Q}}(2S) &=& \frac{6804 \pi^5 T^4}{100 m_Q^2} \frac{(c_2 - \frac{v^2}{2} 
c_4)}{((c_2- \frac{v^2}{2} 
c_4)^2 + (c_1 - \frac{v^2}{2} c_3)^2 )} ~-~ \frac{1035 \sqrt{2} \pi T B_2 }{256 \sqrt{10 m_Q A_2}} 
{(1- v^2)}^{\frac{1}{4}} \frac{(10-5 v^2 -6 v^4)}{24(1-v^2)} \nonumber \\ 
&\times& \sqrt{\frac{\sqrt{(~(c_2- \frac{v^2}{2} 
c_4)^2 + (c_1 - \frac{v^2}{2} c_3)^2 )} - (c_2- \frac{v^2}{2} c_4) }
 {2 (~(c_2- \frac{v^2}{2} c_4)^2 + (c_1 - \frac{v^2}{2} c_3)^2 )}}~,
\end{eqnarray}
where the second term is the higher-order correction in velocity and 
contribute smaller to the first term 
in the limit of small velocities (which is also checked numerically)
and thus can be neglected.
\begin{figure}
\begin{center}$
\begin{array}{cc}
\includegraphics[keepaspectratio,width=3.0in,height=3.0in]{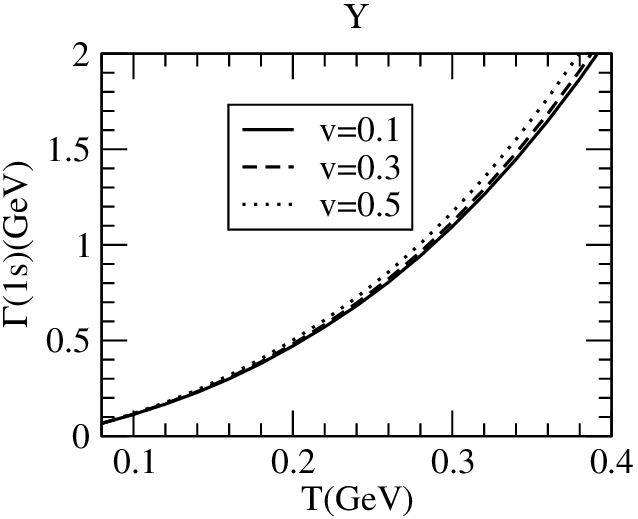} 
\hspace{1.1cm}
\includegraphics[keepaspectratio,width=3.0in,height=3.0in]{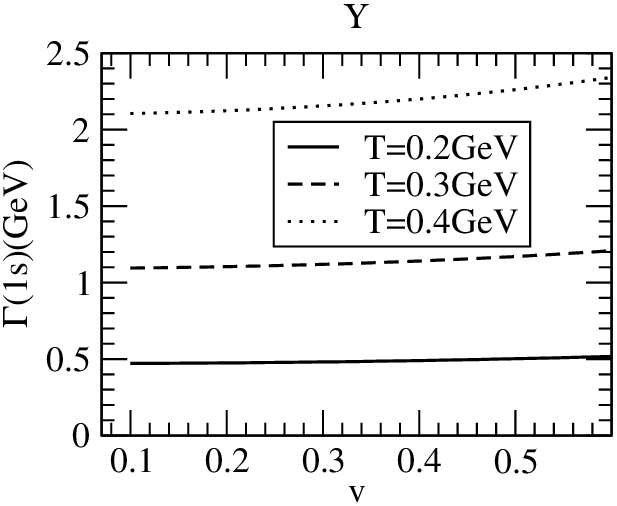}
\end{array}$
\end{center}
\caption{\footnotesize Same as Figure 2 but for the bottomonium state.}
\label{bot1s}
\end{figure}
In the limit of small velocity, the thermal width for 1S state can be 
simplified into
\begin{eqnarray}
\Gamma_{Q\bar{Q}}(1S) \approx  \frac{972 \pi^5 T^4}{200 m_Q^2}\bigg[ \frac{c_2}{(c_1^2 + 
c_2^2)} + \frac{v^2}{2}\left(-\frac{c_4}{(c_1^2 + c_2^2)} + \frac{2 c_2( c_2 c_4 + c_1 
c_3)}{(c_1^2 + c_2^2)^2}   \right) \bigg]
\end{eqnarray}
We have plotted the thermal width for the ground and first excited 
states of charmonium (in Fig.(\ref{charm1s}) and (\ref{charm2s}) ), 
where we found that the width is always broadened with the the increase of 
temperature and also increases slowly with the velocity. It can be 
understood from the fact that as the velocity of the medium increases,
\begin{figure}
\begin{center}$
\begin{array}{cc}
\includegraphics[keepaspectratio,width=3.0in,height=3.0in]{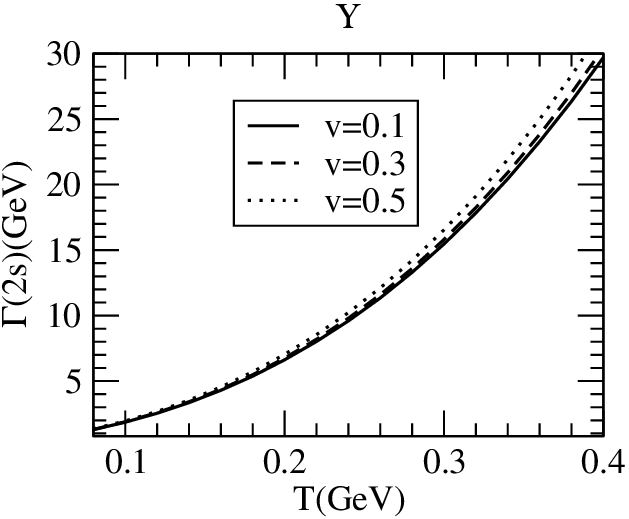}
 \hspace{1.1cm}
\includegraphics[keepaspectratio,width=3.0in,height=3.0in]{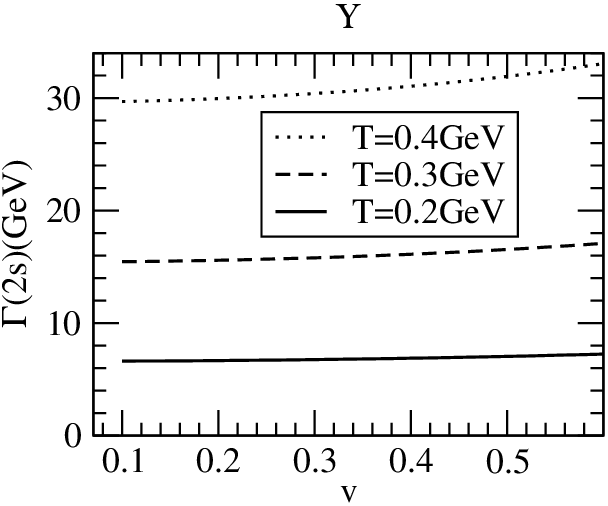}
\end{array}$
\end{center}
\caption{\footnotesize Same as Figure 3 but for the bottomonium state.}
\label{bot2s}
\end{figure}
the flux of plasma passing through the $Q \bar{Q}$ pair increases and hence 
causes to broaden the thermal width. The same reason applies to
the increase of imaginary part of potential with the velocity. 
As a consequence, the pair will be dissolved earlier into a moving medium
compared to the static medium.
Another observation is that the thermal width for bottomonium  states
(Fig.(\ref{bot1s}) and (\ref{bot2s}))
is much smaller than the charmonium states because bottomonium states are 
bound tighter than charmonium states.
\section{Conclusions}
In summary, we have obtained the inter-quark potential at finite temperature 
in a dual gravity closer to thermal QCD for a heavy quark and 
antiquark pair moving normal to its orientation.
When the (hanging) string lying on the fifth 
dimension, $z_{\rm{max}}$ is far above the horizon, ie.
$z_{\rm{max}} << z_h$, the Nambu-Goto action gives rise to the similar
form of Cornell potential, unlike the Coulomb term alone 
usually reported in the literature. On the other hand when the string reaches 
deep into the horizon ($z_{\rm{max}}>>z_h$), the potential develops an imaginary 
component. Alternatively it can be stated that beyond a critical separation of 
$Q\bar Q$ pair, the string coordinates become imaginary and as a
result, the potential becomes complex. The imaginary part vanishes
in the limit of small separation which is also seen in
the perturbative calculations.
Furthermore as the pair starts moving, the screening of the potential
becomes smaller, which may be understood qualitatively by the decrease of 
effective temperature. However for a particular velocity of the pair, the 
screening becomes stronger with the increase of the temperature 
as observed in the potential for a static pair. 

Since the quarkonium dissociation is presently thought mainly due to
Landau damping induced thermal width, we have therefore obtained 
the thermal width for the ground and first excited states from the
imaginary part of the potential. We found that the width 
not only increases with the increase of 
temperature, it also increases slowly with the velocity of the
pair, which makes sense. The effect of enhancement of width is more 
pronounced to the charmonium states than the bottomonium states.
Since the study of quarkonium bound states propagating in the QGP 
at  finite velocity  is nowadays relevant in heavy-ion collisions
so our study gives an theoretical input in this regard.
Indeed the PHENIX Collaboration~\cite{Adare:2006nq} had found a
substantial elliptic flow for heavy-flavor electrons, 
indicating a significant damping of heavy quarks while  
travelling across the medium~\cite{vanHees:2007me}. 
\footnote{{\bf Note added.} When this article was being finished, we became 
aware of Ref.\cite{Noronha:2015JHEP} which also discussed the effects of 
nonzero rapidity on the imaginary part of the heavy quark potential 
by the thermal worldsheet fluctuation method in 
a ${\cal N}$=4 supersymmetric plasma. In 
their work, they computed the imaginary potential for arbitrary orientations 
of the $Q \bar Q$ pair with respect to the hot plasma wind in pure AdS BH
background. On the contrary we use OKS-BH geometry in the gravity side whose
dual in the gauge theory side is closer to thermal QCD {\em i.e.} runs with
the energy scale similar to QCD, unlike scale-independent in
${\cal N}$=4 SYM theory. Moreover both the origin of complex nature of potential
and renormalization subtraction are different from the above 
work~\cite{Noronha:2015JHEP} {\em namely} in our work $z_{\rm{max}}$ 
(the maximum vale of $z$ along the fifth direction, along which the string 
is hanging) becomes complex for a critical separation of the pair and 
leads to the complex-valued string coordinates~\cite{Yavier:PRD2008,Hatsuda:PRD2013}, hence 
the potential too becomes complex.}

\section{Acknowledgements}
BKP is thankful to the Council of Scientific and Industrial Research
project (03 (1215)/12/EMR-II), Government of India for the financial 
assistance.

\end{document}